\documentclass[journal=jacsat,manuscript=article]{achemso}

\usepackage[version=3]{mhchem}
\usepackage{color}

\newcommand{\etal}{\textit{et al.}}

\author{Klichchupong Dabsamut}
\affiliation{Department of Physics, Faculty of Science, Kasetsart University, Chatuchak, Bangkok 10900, Thailand}
\author{Intuon Chatratin}
\affiliation{Department of Materials Science and Engineering, University of Delaware, Newark, Delaware 19716, USA}
\author{Thanasee Thanasarnsurapong}
\affiliation{Department of Physics, Faculty of Science, Kasetsart University, Chatuchak, Bangkok 10900, Thailand}
\author{Tosapol Maluangnont}
\affiliation{Electroceramics Research Laboratory, College of Materials Innovation and Technology, King Mongkut’s Institute of Technology Ladkrabang, Ladkrabang, Bangkok 10520, Thailand}
\author{Adisak Boonchun}
\email{adisak.bo@ku.th}
\affiliation{Department of Physics, Faculty of Science, Kasetsart University, Chatuchak, Bangkok 10900, Thailand}

\title[An \textsf{achemso} demo]
{Theoretically proposed another stable polymorph of two-dimensional \textit{penta}-PdPSe}

\abbreviations{IR,NMR,UV}

\begin{document}
\pagebreak
	
	\begin{abstract}
		The theoretical discovery of new and stable 2D \textit{penta} materials has stimulated the technological advancement due to the anticipated exotic properties of such structure, including the recent $\alpha$ phase and $\beta$ phase of \textit{penta}-NiPS based on first-principle calculations. Inspired by the similarity between the theoretically proposed \textit{penta}-NiPS and the experimentally synthesized ($\alpha$ phase) of \textit{penta}-\ce{PdPSe}, we proposed herein the $\beta$ phase \textit{penta}-PdPSe as a new member of the \textit{penta}-2D materials. Comprehensive analyses indicated that the $\beta$ phase \textit{penta}-\ce{PdPSe} is thermodynamically, dynamically, mechanically, and thermally stable,  similar to the NiPS analog. It was found that $\beta$ \textit{penta}-PdPSe is a wide band gap semiconductor with an indirect band gap of 1.58 eV, significantly lower than 2.15 eV for the $\alpha$ phase. Moreover, the two polymorphs of \textit{penta}-PdPSe are a soft material with 2D Young’s modulus of $E_a$ = 151 Nm$^{-1}$ and $E_b$ = 123 Nm$^{-1}$ for the $\beta$ phase, to be compared with $E_a$ = 155 Nm$^{-1}$ and $E_b$ = 113 Nm$^{-1}$ for the $\alpha$ phase. The calculated absorption coefficient showed that $\beta$ phase \textit{penta}-PdPSe is acceptable for electronic and optical nanodevice.
	\end{abstract}

	\pagebreak
	
	\section{Introduction}
	Since the discovery of graphene, two-dimensional (2D) materials have been steadily progressed in terms of production and materials design alike. Theoretical explorations have expanded the library of stable 2D materials beyond the conventional honeycomb-like structure, including the understanding of composition-structure-properties relationships. Recently, Zhang \textit{et al.}\cite{Zhang2372} theoretically proposed \textit{penta}-graphene with carbon atoms uniquely resembling Cairo pentagonal tiling. Since then, the properties of the \textit{penta}-graphene  have been thoroughly explored through first-principles calculations, for potential applications in e.g., nanoelectronics and nanomechanical devices. Other theoretically proposed examples are \ce{SiC2}, \ce{CN2}, \ce{BN2}, \ce{B2C}, \ce{PdS2}, \ce{AlN2}, \ce{PtN2}, \ce{PdN2}, and \ce{NiS2}. \cite{penta-SiC2,CN2,BN2,B2C,cheng2021pentagonal,penta-PdS2,penta-AlN2,penta-NiS2}. From the experimental point of view, the first example of laboratory-synthesized pentagonal structure is \textit{penta}-\ce{PdSe2} via the mechanical exfoliation of bulk crystals.\cite{pdse2} The \textit{penta}-\ce{PdSe2} based field-effect transistors (FETs) exhibited band gap tunability and high electron mobility.\cite{pdse2} 
	
	The investigation of ternary compounds provides a higher degree of freedom in tuning the properties of \textit{penta}-2D materials \cite{BCN}. The prominent example is 	\textit{penta}-BCN\cite{BCN,dabsamut2021strain} which is thermodynamically, mechanically, and thermally stable according to density functional theory (DFT), in addition to possessing a high piezoelectric response. Note that it is essential that one comprehensively explores various aspects of materials' stability. Later on,  \textit{penta}-\ce{PdPSe} and \textit{penta}-\ce{PdPS} with intrinsic in-plane anisotropy were also proposed and successfully synthesized.\cite{PdPSe,PdPS} (These two examples  are the ternary compound analog of  \textit{penta}-\ce{PdSe2} discussed above.) The \textit{penta}-\ce{PdPSe} and \textit{penta}-\ce{PdPS} are made up from two sublayers which are uncommon in nature, leading to unusual electronic and optical properties.\cite{mortazavi2022,bafekry2022two,PdPSe,PdPS} Recently, Dabsamut \etal \cite{penta-NiPS} reported the two polymorphs of \textit{penta}-NiPS ($\alpha$ and $\beta$ phase) based on first-principle calculations. The $\alpha$ phase has the structure similar to the successfully synthesized \textit{penta}-\ce{PdPSe}. Accordingly, it is interesting to explore if the analogous $\beta$ phase would exist for \textit{penta}-\ce{PdPSe}.
	
	Herein, we employed first-principles calculations and discovered the $\beta$ phase of \textit{penta}-\ce{PdPSe} as another stable polymorph. It is found that The $\beta$ phase \ce{PdPSe} is dynamically, mechanically, and thermally stable, as with the $\beta$ phase \textit{penta}-NiPS.\cite{penta-NiPS} The calculated band gap showed that $\beta$-\ce{PdPSe} is a wide band gap semiconductors with an indirect band gap of 2.18 eV. The Young's modulus and optical response of the two polymorphs are also reported. 
	
	\section{Computational Method}
	We employed density functional theory (DFT) and the projector-augmented wave (PAW) formalism, as implemented in the  Vienna Ab-initio Simulation Package (VASP) code. Perdew-Burke-Ernzerhof (PBE) functional was used and the cutoff energy was set to 500 eV for the plane-wave expansion. For a unit cell, the structure geometries were fully optimized until total Hellmann-Feynman forces on every atoms were less than 0.02 eV/\AA and the energy convergence was set to $10^{-6}$ eV. The Brillouin zone was sampled with a $7 \times 7 \times 1$ k-point mesh within the Monkhorst-Pack scheme. To avoid the interaction between a layer and its periodic images, a vacuum region of 16 \AA was introduced along the direction perpendicular to the layer. As the standard DFT severely underestimate band gaps, we used the Heyd-Scuseria-Ernzerhof (HSE06) hybrid functional to describe electronic structure of the materials. Phonon spectra were calculated based on density-functional perturbation theory (DFPT) method, as implemented in a PHONOPY package. The calculations were performed in a $3 \times 3 \times 1$ supercell, and energy cutoff of 500 eV and a single k-point at $\Gamma$ were used. To confirm thermodynamic stability, we performed ab initio molecular dynamics (AIMD) simulation.

	\section{Results and discussion}
\begin{figure}[h]
	\begin{center}
	    \includegraphics[width=9 cm]{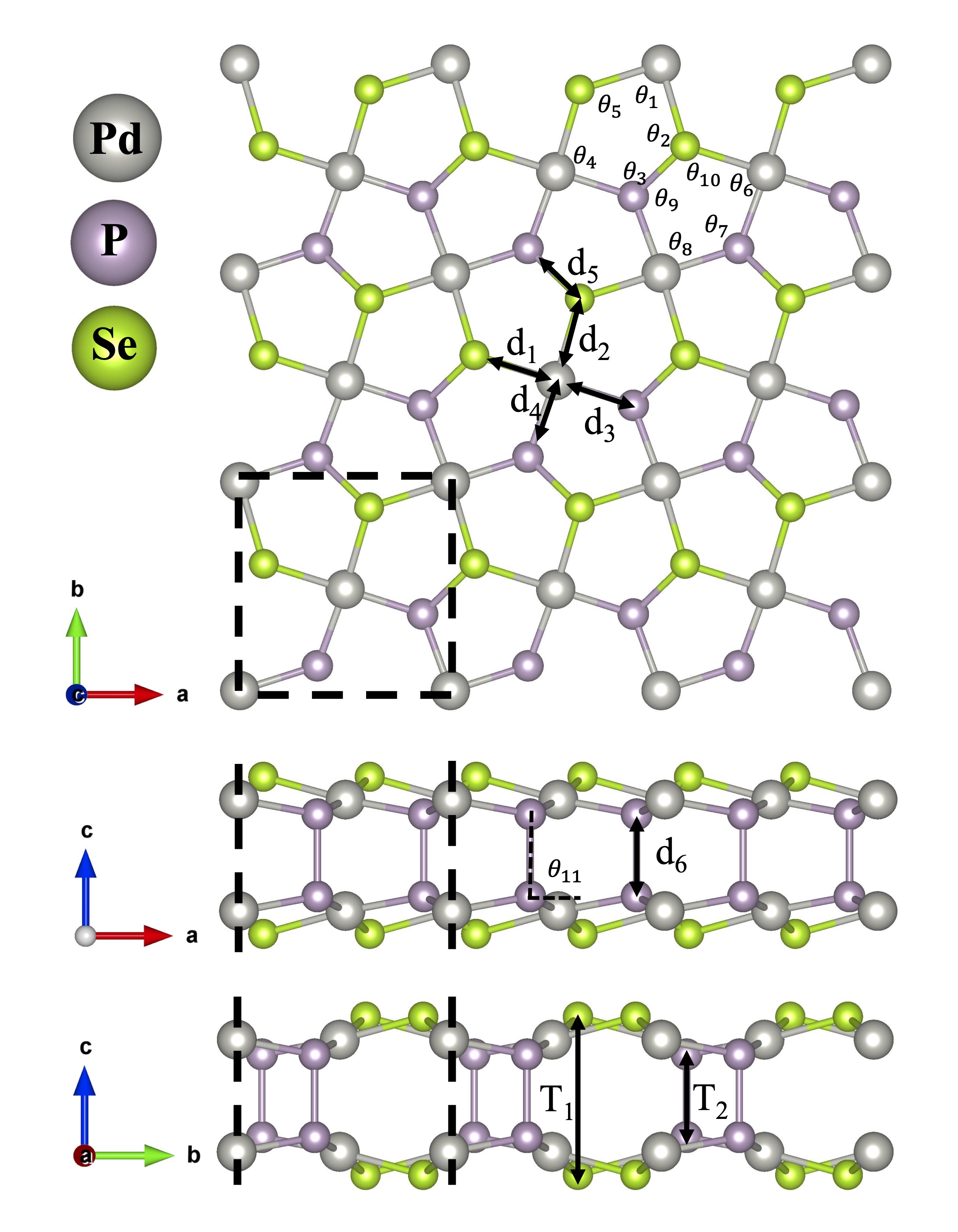} 
	    \caption{\ 3x3 supercell of pentagonal $\beta$-PdPSe geometrical structures.}
	\label{fig1}
	\end{center}
    \end{figure}

	We first constructed the \textit{penta}-PdPSe monolayer based on the reported atomic configuration of the new polymorph, $\beta$ NiPS\cite{penta-NiPS}, by replacing the Ni atoms with Pd atoms and S atom with Se atoms. This is named the $\beta$-phase composing of two sublayers linking by the -(P-P)- bonding as shown by the side view of Fig. \ref{fig1}. Within one sublayer, every Pd atom bonded to two P atoms and to two Se atoms, showing the pentagon structure, as shown in the top view of Fig. \ref{fig1}. The pentagonal primitive unit cell, shown as the square in Fig. \ref{fig1}, consists of 12 atoms including 4 Pd atoms, 4 P atoms, and 4 Se atoms. The lattice constants of the $\beta$ phase \textit{penta}-PdPse  are \textit{\textbf{a}} = 5.89 \AA\ and \textit{\textbf{b}} = 5.84 \AA. The whole thickness of the layer, T$_1$, is 4.35 \AA, while the distance between two sublayers, T$_2$, is 2.27 \AA\ .  
	
In addition to the presently proposed $\beta$ phase, we constructed the $\alpha$-PdPSe based on the reported atomic configuration. Our calculated lattice constants of the $\alpha$ phase  are \textit{\textbf{a}} = 5.86 \AA\ and \textit{\textbf{b}} = 5.79 \AA, exactly matching the previously reported values.\cite{bafekry2022two} The whole thickness T$_1$ of $\alpha$ phase is 4.41 \AA, while the sublayer distance T$_2$ is 2.17 \AA. The larger T$_2$ of the $\beta$ phase is understandable because the direct stacking generates a higher degree of coulomb repulsion of the same atoms between two sublayers, as is commonly observed in other 2D materials. The fundamental properties of the two phases including bond lengths, bond angles and band gap are listed in Table. \ref{tbl:lattice}. 

\begin{table*}
\small
  \caption{\ The calculated lattice constants \textit{\textbf{a}} and \textit{\textbf{b}} in \AA, bond lengths $d_1$-$d_6$ in \AA, bond angles $\theta_1$-$\theta_{11}$ in degree, thicknesses T$_1$ and T$_2$ in \AA, the cohesive energy per atom in eV/atom, and band gap $E_g^{PBE}$ and $E_g^{HSE}$ in eV}
  \label{tbl:lattice}
  \begin{tabular*}{\textwidth}{@{\extracolsep{\fill}}llllllllll}
    \hline
	& \textit{\textbf{a}} & \textit{\textbf{b}} & d$_1$ & d$_2$ & d$_3$ & d$_4$ & d$_5$ & d$_6$ &  \\ \hline
	$\alpha$-PdPSe & 5.86 & 5.79 & 2.48 & 2.47 & 2.31 & 2.30 & 2.29 & 2.20 &   \\
	$\beta$-PdPSe & 5.89 & 5.84 & 2.46 & 2.47 & 2.31 & 2.31 & 2.27 & 2.27 &    \\ \hline
	& $\theta_1$ & $\theta_2$ & $\theta_3$ & $\theta_4$ & $\theta_5$ & $\theta_6$ & $\theta_7$ & $\theta_8$ & $\theta_9$  \\ \hline
	$\alpha$-PdPSe & 84.68 & 103.82 & 109.01 & 94.02 & 114.86 & 90.25 & 123.65 & 90.46 & 104.78 \\
	$\beta$-PdPSe & 85.23 & 106.52 & 109.50 & 93.89 & 116.95 & 90.14 & 124.15 & 90.45 & 106.13   \\ \hline
	&  $\theta_{10}$ & $\theta_{11}$ & T$_1$ & T$_2$ & $E_{coh}$ &   $E_g^{HSE}$ & &    &  \\ \hline
	$\alpha$-PdPSe & 105.55 & 81.07 & 4.42 & 2.17 & -4.80 &  2.15 &  & &   \\
	$\beta$-PdPSe & 107.04 & 90.00 & 4.35 & 2.27 & -4.76 & 1.58 &  & &    \\ 
    \hline
  \end{tabular*}
\end{table*}

\begin{figure}
	\begin{center}
	\includegraphics[width=12cm]{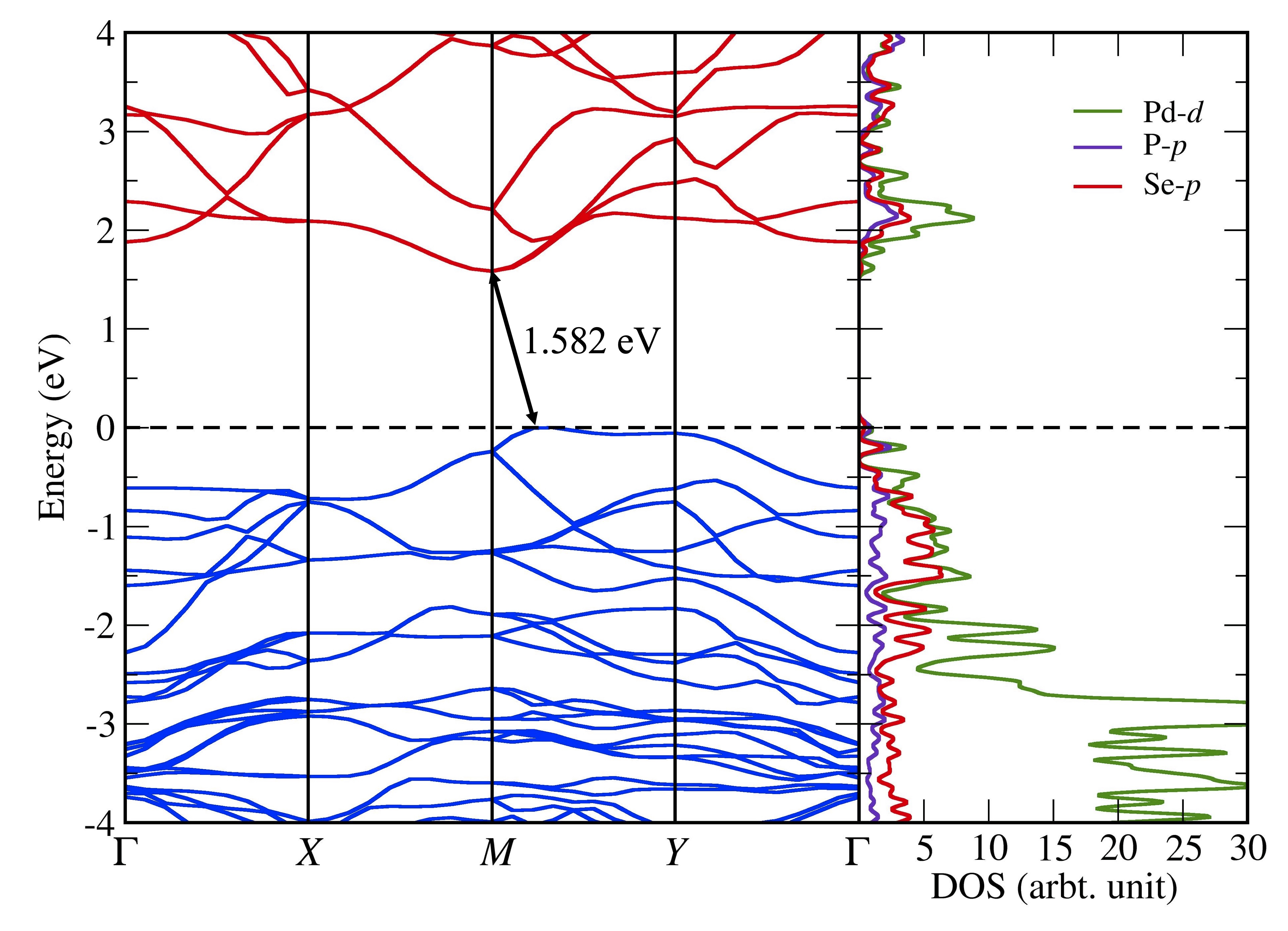} 
	\caption{\ Calculated band structure and partial density of state (PDOS) for $\beta$-PdPSe.}
	\label{fig:banddos}
	\end{center}
	\end{figure}

	The electronic properties of the $\beta$-PdPSe phases are first elucidated by the calculated band structure using the HSE06 functional as shown in Fig.\ref{fig:banddos}. The electronic band structures show the indirect band gaps of 1.58 eV for the $\beta$ phase, significantly larger than 2.15 eV for the $\alpha$ phase. As seen in  Fig.\ref{fig:banddos} within HSE06, the valence band maximum (VBM) of $\beta$ phase is located along the $M-Y$ path and the conduction band minimum (CBM) at the $M$ point.
	
	To further analyze the electronic properties of \textit{penta}-PdPSe, the partial density of states (PDOS) obtained from the HSE06 functional were plotted along with the band structure in Fig.\ref{fig:banddos}. For $\beta$-PdPSe, both of VBM and CBM are mainly contributed by the Pd-$d$ states. The larger indirect band gap in the $\alpha$ phase compare to the $\beta$ phase is similar to the previous study of \textit{penta}-NiPS.\cite{penta-NiPS}

	To evaluate the crystal binding stability, the cohesive energy per atom was calculated, given by the following equation:
	\begin{equation}
	E_{coh} = \frac{E_{tot}(\text{PdPSe})-\sum n_i E_i}{\sum n_i}
	\end{equation}
	where $E_{tot}(\text{PdPSe})$ is the total energy of \textit{penta}-PdPSe, $n_i$ is the number of Pd, P and Se atoms in the unit cell and $E_i$ represents the energies of an isolated single Pd, P and Se atom. In the present work, the calculated cohesive energy of \textit{penta}-PdPse are -4.80 and -4.76 eV/atom for the $\alpha$ and $\beta$ phase, respectively. The negative values indicate that both polymorphs of \textit{penta}-PdPSe are more energetically favorable than the isolated atoms. In addition, the $\alpha$ phase is just slightly more energetically favorable than the $\beta$ phase. The slight difference suggests that they could potentially be synthesized by a fine tuning of synthetic conditions such as temperature, pressure, surrounding gas, among others (as is the case with more common solids such as anatase and rutile $\rm TiO_2$.\cite{cohesiveTiO2_p1,cohesiveTiO2_p2}  Note that the synthesise of $\alpha$  \textit{penta}-PdPSe has been experimentally accomplished, \cite{PdPSe} but the selective synthesis of the $\beta$ phase remains undone. 
	
	\begin{figure}[h]
	\begin{center}
	\includegraphics[width=10cm]{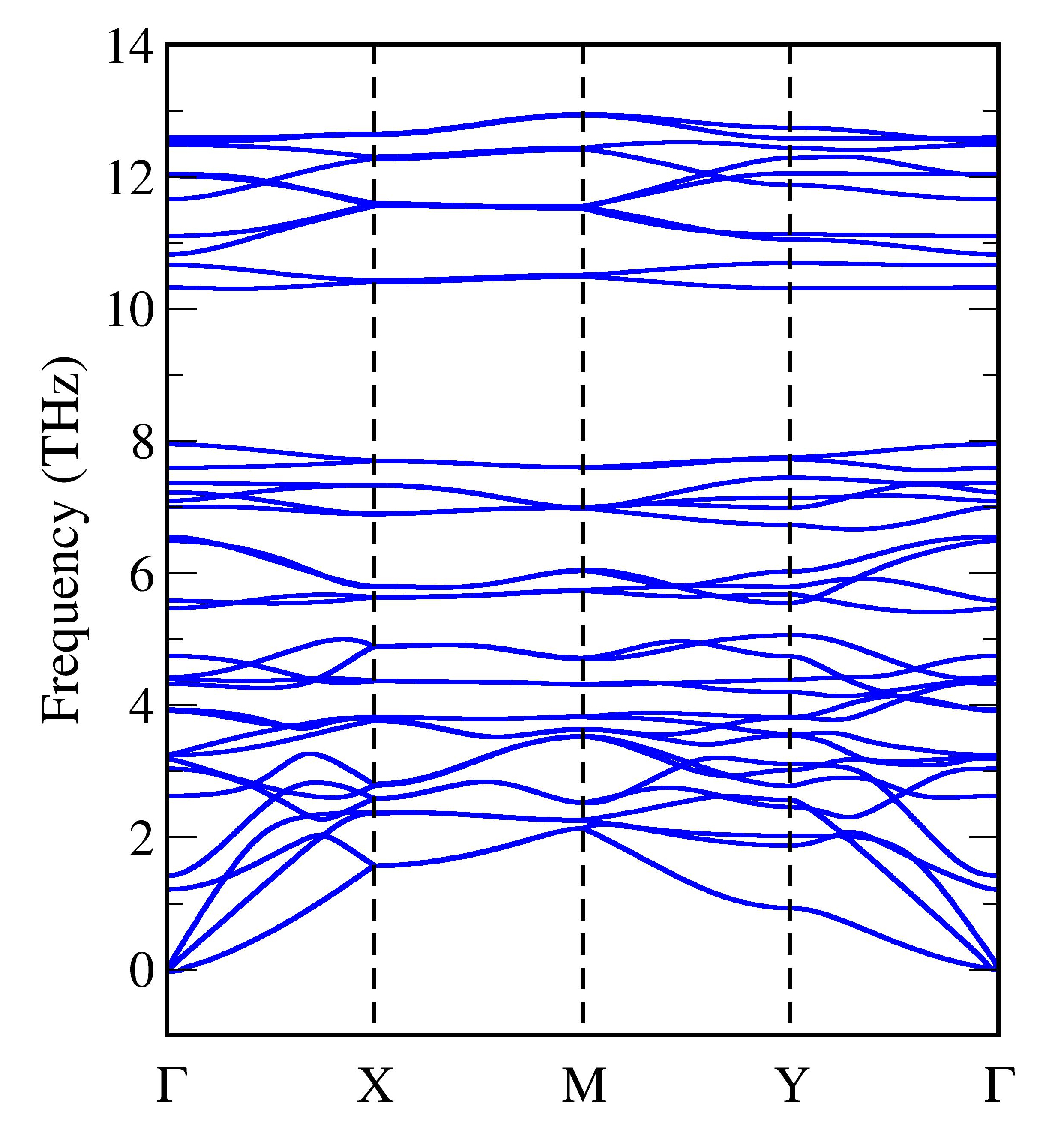} 
	\caption{\ Calculated phonon band structure of $\beta$-PdPSe.}
	\label{fig2}
	\end{center}
	\end{figure}

To further verify the dynamic stability of $\beta$ \textit{penta}-PdPSe, we calculated the phonon dispersion. The dynamic stability is indicated by the positive frequencies in the phonon spectra throughout the entire  Brillouin zone. Meanwhile, the negative values are referred to as the the imaginary frequency and subsequently dynamical instability. Fig \ref{fig2} shows the calculated phonon dispersion of $\beta$ \textit{penta}-PdPSe. There is no imaginary modes throughout the Brillouin zone, confirming its dynamic stability. (The dynamical stability of $\alpha$ \textit{penta}-PdPSe has been validate in the previous work by others.) \cite{pdse2}
	
In addition to dynamic stability, we also verified the mechanical stability of both polymorphs of \textit{penta}-PdPSe. This can be accomplished by considering the linear elastic constants, C$_{11}$, C$_{22}$, C$_{12}$ and C$_{66}$, which are elements in the stiffness tensor. Here, the elastic constants are obtained by finite difference method as listed in Table 2. (The elastic constants $C_{jk}$ of $\alpha$-PdPSe has not yet been reported.) Typically, a structure is considered mechanically stable when the linear elastic constants satisfy the conditions C$_{11}$C$_{22}$-C$_{12}^2 > 0$ and C$_{66} > 0$. Notably, these conditions are obeyed for both phases, confirming their mechanical stability.  
\begin{table}[h]
\small
  \caption{\ The elastic constants $C_{jk}$ (Nm$^{-1}$), Poisson's ratio on [100] ($\nu_a$) and [010] ($\nu_b$) directions and Young's modulus (Nm$^{-1}$) on [100] ($E_a$) and [010] ($E_b$) directions of $\alpha$-PdPSe and  $\beta$-PdPSe.}
  \label{tbl:constant}
  \begin{tabular*}{0.48\textwidth}{@{\extracolsep{\fill}}lll}
    \hline
	& $\alpha$-PdPSe & $\beta$-PdPSe \\ 
    \hline
	$C_{11}$ & 157.58 & 154.49  \\
	$C_{12}$ & 18.55 & 22.82  \\
	$C_{22}$ & 115.67 & 126.83  \\
	$C_{66}$ & 44.04 & 43.99   \\
	$\nu_a$  & 0.118 & 0.148  \\
	$\nu_b$ & 0.160 & 0.180   \\
	$E_a$ & 155.40 & 151.12   \\
	$E_b$ & 112.70 & 122.73   \\
    \hline
  \end{tabular*}
\end{table}

Next, we evaluated the 2D Young’s modulus on the [100] and [010] directions (i.e., the in plane  direction), which is obtained by $E_a =$(C$^2_{11}$ - C$^2_{12}$)/C$_{11}$ and $E_b =$ (C$^2_{22}$ - C$^2_{12}$)/C$_{22}$. In the
$\beta$ phase, $E_a$ = 151.12 Nm$^{-1}$ and $E_b$ = 122.73 Nm$^{-1}$. Meanwhile, in the 
$\alpha$ phase, $E_a$ = 155.40 Nm$^{-1}$ and $E_b$ = 112.70 Nm$^{-1}$. Accordingly, the  $\beta$ phase is harder than the $\alpha$ phase in [010] direction. In addition, Table \ref{tbl:constant} lists the calculated Poisson’s ratio. For $\beta$-PdPSe, the calculated Poisson’s ratio on the [100] direction, $\nu_a$ = C$_{12}$/C$_{11}$ is 0.148 which is lower than that on the the [010] direction, $\nu_b$ = C$_{12}$/C$_{22}$ of 0.180. For $\alpha$-PdPSe, the calculated Poisson’s ratios are $\nu_a$ = 0.118 and $\nu_b$ = 0.160, respectively. 
	
\begin{figure}[h]
	\begin{center}
	\includegraphics[width=14cm]{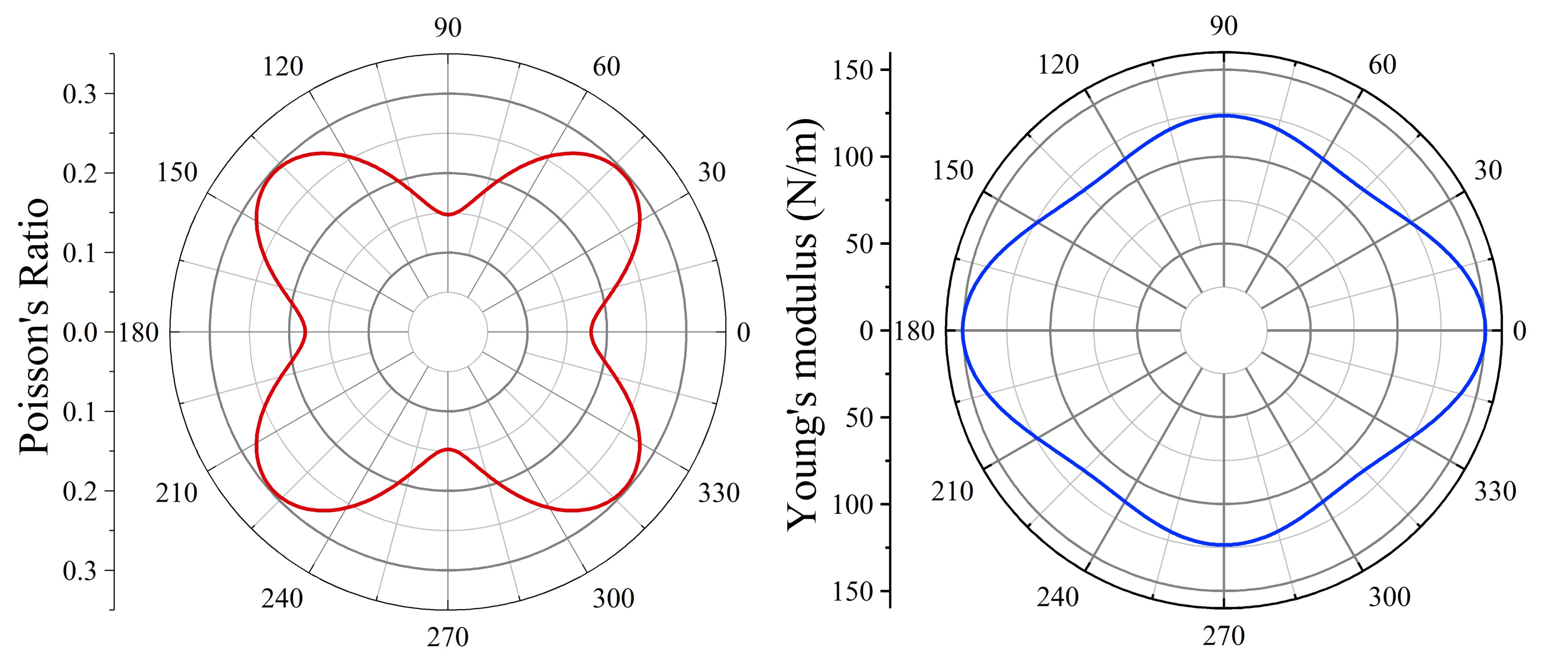} 
	\caption{\  Variation of (a) in-plane Young's modulus,  and (b) Poisson's ratio for $\beta$-PdPSe.}
	\label{fgr:polar}
	\end{center}
\end{figure}
	
	To further explore the dependence of mechanical properties on the crystal orientation of $\beta$-PdPSe, we computed the Young's modulus ($E$) and Poisson's ratio ($\nu$) along an arbitrary direction $\theta$ ($\theta$ is the angle relative to the x direction) using the following formulas:
	
	\begin{eqnarray}
	E(\theta) = \frac{C_{11}C_{12}-C_{12}^2}{C_{11}s^4+C_{22}c^4+\left(\frac{C_{11}C_{12}-C_{12}^2}{C_{66}}-2C_{12}\right)c^2 s^2} \\ [3pt]
	\nu(\theta) = \frac{\left(C_{11}+C_{12}-\frac{C_{11}C_{12}-C_{12}^2}{C_{66}}\right)c^2 s^2-C_{12}(c^4+s^4)} {C_{11}s^4\theta+C_{22}c^4+\left(\frac{C_{11}C_{12}-C_{12}^2}{C_{66}}-2C_{12}\right)c^2 s^2} 
	\end{eqnarray}
	where $c=\cos\theta$ and $s=\sin\theta$. The calculated results are plotted in Fig \ref{fgr:polar}. One can see that the maxima of the in-plane Young’s modulus for $\beta$-PdPSe are located at the [100] direction with the value of 150.38 ${\rm Nm^{-1}}$. Meanwhile, the minima are  at $\sim 53^{\circ}$ with the value of 112.79 ${\rm Nm^{-1}}$. Regarding the Poisson’s ratios, the minimum are located at $90^{\circ}$ with the values of 0.148, while the maximum is located at $\sim 43^{\circ}$ with values of 0.297.  Because the in-plane Young’s modulus and Poisson’s ratios are dependent on crystal orientation, $\beta$-PdPSe \textit{penta}-PdPSe exhibits anisotropic mechanical properties which can be explained by its lattice constant where $a\neq b$. 
	
\begin{figure}[h]
	\begin{center}
	\includegraphics[width=14cm]{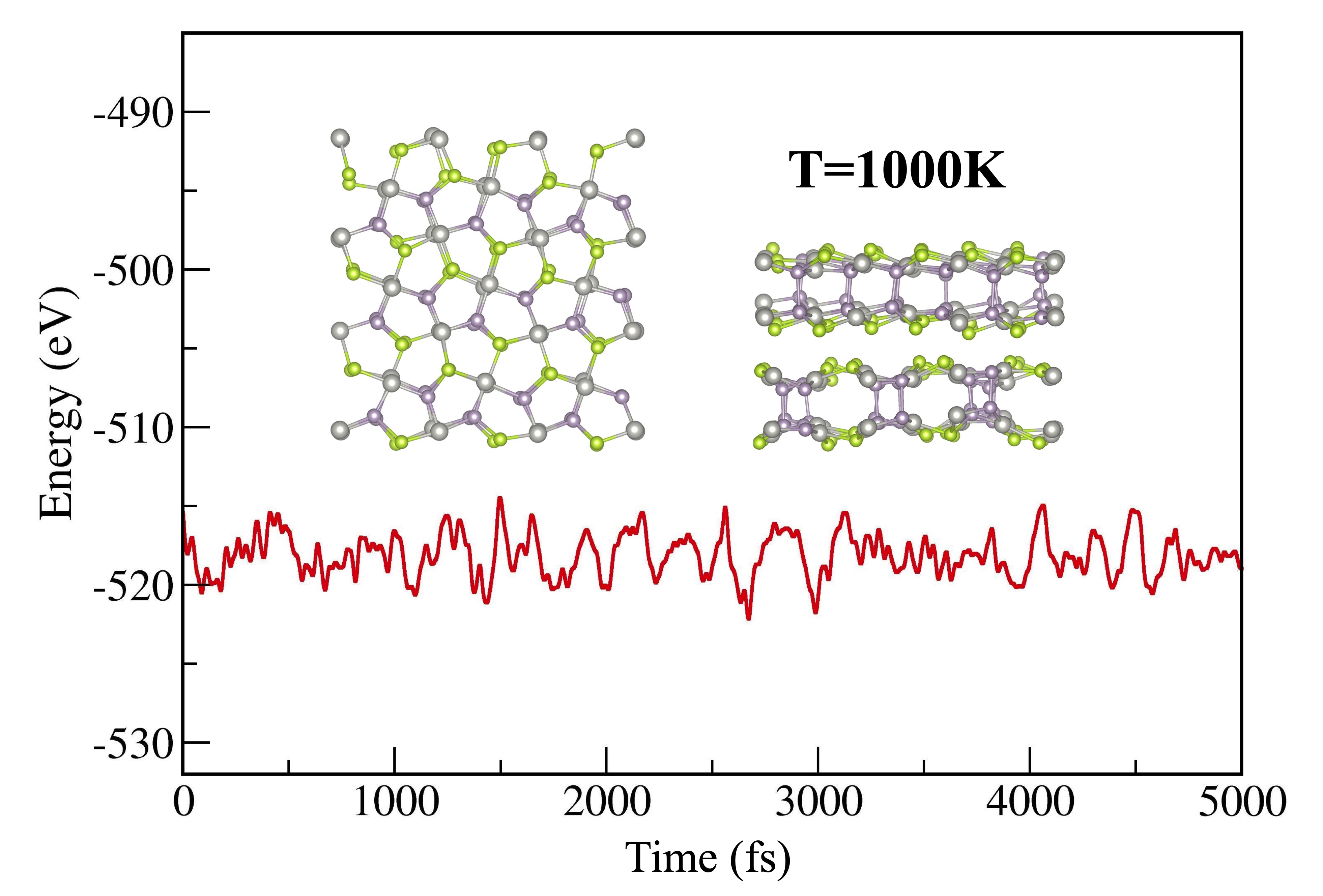} 
	\caption{\ Total energy fluctuation with time during the AIMD simulation at 1000 K of $\beta$ phase \textit{penta}-PdPSe.}
	\label{fig3}
	\end{center}
\end{figure}
	
Next, the thermal stability of \textit{penta}-PdPSe was examined via ab initio molecular dynamic simulations. Supercells with a duplication of a $4 \times 4 \times 1$ of the unit cell is used. The molecular dynamics simulations were performed for 5 ps with a time step of 1 fs under the temperature at 1000 K. Fig. \ref{fig3} shows the fluctuation of the potential energy for $\beta$ \textit{penta}-PdPSe throughout the simulation time at 1000 K. It is found that the $\beta$ structure reaches the steady state and equilibrates around a constant energy after 1,000 fs, indicating its thermal stability at least up to 1000 K.

\begin{figure}[h]
	\begin{center}
	\includegraphics[width=14cm]{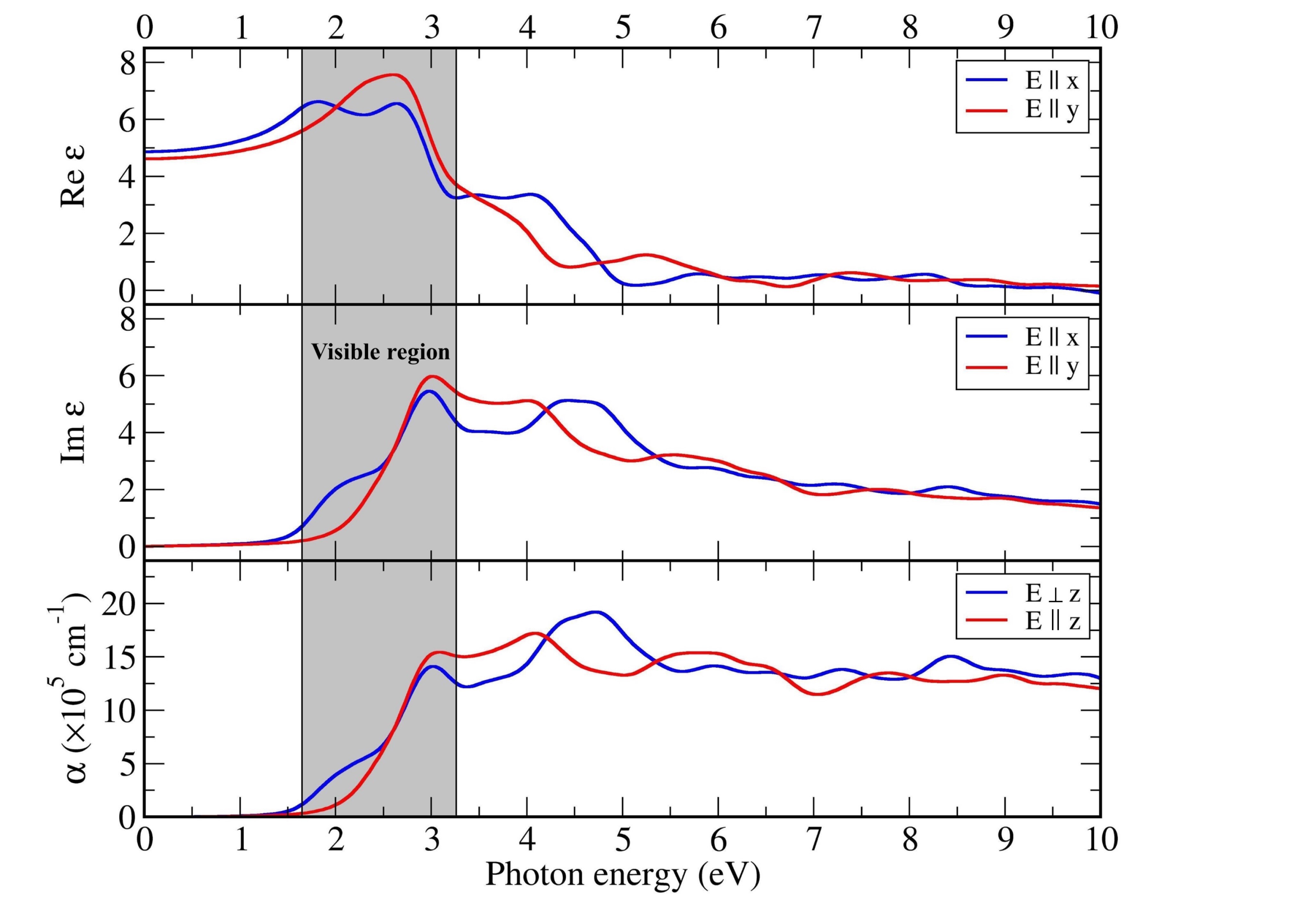} 
	\caption{\ Real part and Imaginary part of the dielectric function as a function of in-plane photon energy (x- and y-polarized light) for $\beta$ phase \textit{penta}-PdPSe monolayers.}
	\label{fgr:optical}
	\end{center}
\end{figure}

Furthermore, we explore the optical response of the $\beta$-PdPSe using the RPA method constructed over HSE06 results with $21\times21\times1$ k-point grid. Due to the uneven geometry along the x- and y- axes, the two optical spectra are anisotropic for light polarization along the in-plane directions, i.e. the x- and y- polarized directions (E$\parallel$x and E$\parallel$y). The optical spectra in the perpendicular direction are disregarded due to the fact that depolarization contribution is huge.\cite{saleem2017investigation,mortazavi2022} The calculated  of dielectric function ($\varepsilon$) in real part (${\rm Re}\; \varepsilon$) and imaginary part (${\rm Im}\;\varepsilon$) in polarized directions \textit{versus} photon energy are depicted in Fig. \ref{fgr:optical}. The absorption edges of imaginary part occur at the energy of $\sim1.47$ and $\sim2.11$ eV  for in-plane direction.  The values of the static dielectric constant (the values of real part of the dielectric function at zero energy) for the novel $\beta$ phase \textit{penta}-PdPSe were measured to be $\sim4.86$ and $\sim4.61$ eV along E$\parallel$x and E$\parallel$y. According to uneven geometry along the x- and y- axes, static dielectric constant along E$\parallel$x and E$\parallel$y are not equal. 

The absorption coefficient along the in-plane (E$\perp$z) and out-of-plane (E$\parallel$z) polarization as a function of the photon energy and wavelength is shown in Fig. \ref{fgr:optical}(c). The absorption edge for the $\beta$ phase \textit{penta}-PdPSe occurs at 1.51 and 2.12 eV or a wavelength of $\sim820$ and $\sim580$ nm along in-plane (E$\perp$z) and out-of-plane (E$\parallel$z), respectively. The absorption coefficients of out-of-plane axes is in the visible light regime and are attained $\sim 10^5\; {\rm cm}^{-1}$, which are comparable to those of organic perovskite solar cells.\cite{jeon2014solvent,shirayama2016optical} These results indicate that the $\beta$  \textit{penta}-PdPSe possesses significant light-harvesting capabilities for the solar spectrum in the visible regime of light. Moreover, the anisotropic optical properties along the out-of-plane directions suggest that they exhibit attractive prospects for the design of electronic and optical nanodevices. 

\section{Conclusions}
By using DFT calculation, we theoretically discovered $\beta$ phase \textit{penta}-PdPSe as the new member in the \textit{penta}-2D family. Therefore, the \textit{penta}-PdPSe is stable in two polymorphs, the successfully synthesized $\alpha$ phase and the presently reported $\beta$ phase. The $\beta$ \textit{penta}-PdPSe is dynamically, mechanically, and thermally stable, as comprehensively verified from  phonon dispersion, elastic constants, and molecular dynamic simulation, respectively. The $\beta$-PdPSe is a semiconductor with an indirect band gap of 2.18 eV. Both $\alpha$ and $\beta$ phases are soft materials with 2D Young’s modulus of $E_a$ = $E_a$ = 155 Nm$^{-1}$ and $E_b$ = 113 Nm$^{-1}$ for the $\alpha$ phase and $E_a$ = 151 Nm$^{-1}$ and $E_b$ = 123 Nm$^{-1}$ for the $\beta$ phase. The absorption coefficient  showed that $\beta$ phase \textit{penta}-PdPSe is acceptable for electronic and optical nanodevice.

\section*{Acknowledgements}
This project was funded by National Research Council of Thailand (NRCT) and Kasetsart University : N42A650278 and was partially funded by the National Research Council of Thailand (Zero Waste Project). T.T. was supported by the Graduate School Fellowship Program from Kasetsart University.  We wish to thank NSTDA Supercomputer Center (ThaiSC) for providing computing resources for this work.

	\bibliography{rsc}
	
\end{document}